The Neural Basis and Evolution of Divergent and Convergent Thought

Liane Gabora

University of British Columbia

Note: This is the author's pre-publication version of the manuscript. It may have minor deviations from the final, published version.

RUNNING HEAD: *[Neural Basis Evolution Divergent Convergent]*

Correspondence should be addressed to:

Liane Gabora

Department of Psychology, University of British Columbia

Okanagan Campus, Fipke Centre for Innovative Research, 3247 University Way

Kelowna BC, Canada V1V 1V7

Email: liane.gabora[at]ubc.ca

Tel: 250-807-9849

Abstract

This chapter takes as its departure point a neural level theory of insight that arose from studies of the sparse, distributed, content-addressable architecture of associative memory. It is argued that convergent thought is most fruitfully characterized in terms of, not the generation of a single correct solution (as it is conventionally construed), but using concepts in their most compact form by activating neural cell assemblies that respond to their most typical properties. This allows them to be deployed in a conventional manner such that effort is reserved for exploring causal relationships. Conversely, it is argued that divergent thought is most fruitfully characterized in terms of, not the generation of multiple solutions (as it is conventionally construed), but using concepts in a form that is, albeit expanded, constrained by the situation, by activating neural cell assemblies that respond to context-specific atypical properties. This allows them to be deployed in a manner that is conducive to exploring unconventional yet potentially relevant associations, and unearthing potentially useful relationships of correlation. Thus, divergent thought can involve as few as one idea. This proposal is compatible with widespread beliefs that (1) most creative tasks require not many solutions but one, yet entail both divergent and convergent thinking, and (2) not all problems with multiple solutions require creative thinking, and conversely, some problems with single solution *do* require creative thought. The chapter discusses how the ability to shift between convergent and divergent modes of thought may have evolved, and it concludes with educational and vocational implications.

The Neural Basis and Evolution of Divergent and Convergent Thought

It is standard to define convergent thinking problems as those with only one correct solution and divergent thinking problems as open-ended problems with multiple solutions (Guilford, 1950; Runco, 2010), and to essentially equate creativity with the capacity for convergent and divergent thinking (e.g., Onarheim & Friis-Olivarius, 2013), with divergent thinking the more important of the two for creativity. However, this conception of divergent and convergent thinking does not align with what people think of as creative.

First, not all tasks with multiple solutions require creative thinking, such as task of responding to the question "what is something red?" Indeed, some problems with just one correct solution, such as the Remote Associates Test (RAT), require more creativity than coming up with things that are red. Second, it is often said that creative problem solving tasks "require both convergent and divergent thinking," (e.g., Biggs, Fitzgerald, & Atkinson, 1971; Riddell et al, 2007). However, given the conventional view of convergent/divergent thinking problems this makes no sense, for a problem either has one solution or it has multiple solutions; it can not have both. Third, and perhaps most importantly, does it really make sense to define creative thought in terms of the number of correct solutions, as opposed to the process by which those solutions are generated?

In this chapter I hope to convince you that by looking to the neural level mechanisms underlying creative thought, and how these mechanisms evolved, we are led to a new conception of convergent and divergent thought. Specifically, I will argue that convergent thought is characterized, not by the generation of a single correct solution, but by conceiving of concepts in their conventional contexts. Likewise, it will be argued that divergent thought is characterized, not by the generation of multiple solutions, but by playing with 'halo of potentiality' surrounding concepts—new affordances yielded by new contexts—to hone as few as a single idea.

The chapter begins by summarizing a neural level explanation for creative insight that arose from studies of the architecture of associative memory. It then outlines problems with conventional conceptions of divergent and convergent thought. Finally, it proposes a new conception of divergent and convergent thought that is consistent with the architecture of associative memory, evolutionary considerations, and empirical studies of divergent and convergent processing.

## Conventional Conceptions of Divergent and Convergent Thought

Psychological theories of creativity typically involve two modes of thought. As mentioned in the introduction, *divergent thought* is defined and measured in terms of the generation of multiple discrete, often unconventional possibilities, possibly due to defocused attention and facilititatation of free-association by reduced latent inhibition (Carson, Peterson, & Higgins, 2003). It is characterized as intuitive and reflective, and thought to predominate during idea generation (for a review see Runco, 2010; for comparison between divergent/convergent processes in creativity and dual process models of cognition, see Sowden, Pringle, & Gabora, 2015).

Divergent thought is contrasted with *convergent thought,* which as mentioned above is defined and measured in terms of the capacity to perform on tasks where there is a single correct solution. Convergent thought is characterized as critical and evaluative, and it is believed to predominate during the refinement, implementation, and testing of an idea, as involving selection or tweaking of the most promising possibilities.

Note the difference between how divergent and convergent are *defined and measured*, in terms of number of generated and/or acceptable solutions, versus how they are commonly *conceived*, in terms of cognitive processes. If one were to schematically illustrate divergent thought as it is defined and measured, one might draw a set of multiple blobs with well-defined edges to represent the multiple final outcomes generated, whereas if one were to schematically illustrate divergent thought as it is commonly conceived, one might draw a single blob with blurry edges, as is consistent with the literal meaning of divergent as "spreading out" (as in a divergence of a beam of light).[1] Indeed there are several incompatibilities between how we define and measure divergent thought and how we informally conceive of it. The generation of multiple solutions may or may not involve defocused attention and associative memory. Moreover, in divergent thought as it is generally construed, the associative horizons simply widen generically instead of in a way that is tailored to the situation or context, as illustrated in Figure 1. (For these reasons, the term *associative thought* has sometimes been used to refer to creative thinking that involves defocused attention and context-sensitive associative processes, and the term *analytic thought* has sometimes been used to refer to creative thinking that involves focused attention and executive processes.) In short, the way we define and measure divergent and convergent thought is not consistent with how we appear to think about them.

--------------------------------

Insert Figure 1 About Here

--------------------------------

Brief Summary of Research on the Neural Mechanisms of Creative Thinking

We now review some well-established features of associative memory, and how they are believed to come together in creative thinking (Gabora, 2001, 2010; Gabora & Ranjan, 2013).

First, memory is *sparse:* the total number of neurons in the brain is smaller than the total number of stimuli in the world that could potentially be encoded in memory. Therefore, there exist stimuli that no neuron is tuned to respond to. However, as illustrated in Figure 2, the brain is able to encode (and respond to) these stimuli nevertheless because their representation is *distributed*—or spread out across a cell assembly containing many neurons—and likewise each neuron participates in encoding of many items. Neurons exhibit *coarse coding:* although each neuron responds maximally to a particular feature, dimension, or property, it responds to a lesser degree to similar properties. As illustrated in Figure 3, memory is also *content addressable;* there is a systematic relationship between stimulus content and the cell assemblies that encode it, such that memory items are evoked by stimuli that are similar or 'resonant' (Hebb, 1949).

---------------------------------

Insert Figure 2 About Here

---------------------------------

---------------------------------

Insert Figure 3 About Here

---------------------------------

The fact that representations are distributed across cell assemblies of content-addressable neurons that are sensitive to particular high-level or low-level properties ensures the feasibility of forging associations amongst items that are related, perhaps in a surprising but useful or appealing way. This enables reinterpretation of higher-order relations between perceptual stimuli through synchronization of prefrontal neural populations (Penn, Holyoak, & Povinelli, 2008).

Figure 4 provides a simplified illustrative example of how this works. Let us consider the task of inventing a casual chair that would appeal to the free-spirited mindset of the 1960s.[2] We imagine that the designer had recently thrown beanbags with a toddler, and consider what was going on at the neural level during the invention of the beanbag chair. The context of wanting to invent a comfortable chair could have elicited context-driven expansion of the concept CHAIR such that, not just neurons that respond to typical chair properties—such as 'flat surface to sit on'—were activated, but also neurons that respond to context-relevant properties such as 'conforms to shape'. Activation of the neuron that responds to 'conforms to shape' causes associative retrieval of previously encountered items with this property, such as beanbags. The designer recognizes that while a little beanbag conforms to the hand, a giant one might conform to the body. The overlap in the distributed representations of 'CHAIR in the context *comfortable*' and 'BEANBAG' means that there is a route by which the first could elicit associative retrieval of the second, culminating in invention of BEANBAG CHAIR.

--------------------------------

Insert Figure 4 About Here

--------------------------------

Note that an associative memory that encodes items in less detail might not contain a neuron that responds to objects with the property 'conforms to shape'. In this case, the context 'comfortable' could not elicit associative retrieval of BEANBAG and bring about the invention of BEANBAG CHAIR.

Thus, the sparse, distributed, content-addressable nature of memory is critical for creativity. The fact that associations come to mind spontaneously as a result of representational overlap due to sharing of features encoded by content-addressable neurons means there is no need for memory to be searched or randomly sampled for creative associations to be made (Gabora, 2001, 2010). The more detail with which stimuli and experiences are encoded in memory, the greater the degree to which their distributed representations overlap, and the more potential routes by which they can act as contexts for one another and combine. They may have been encoded at different times, under different circumstances, and the relationship between them never explicitly noticed, but some situation could come along and make their relationship apparent.

There is empirical evidence for oscillations in convergent and divergent thinking, and a relationship between divergent thinking and chaos (Guastello, 1998). The capacity to shift between different modes of thought has been referred to as *contextual focus* (CF) (Gabora, 2003). While some dual processing theories (e.g., Evans, 2003) make the split between automatic and deliberate processes, CF makes the split between an associative mode conducive to detecting relationships of correlation, and an analytic mode conducive to detecting relationships of causation. Defocusing attention facilitates associative thought by diffusely activating a broad region of memory, enabling obscure (though potentially relevant) aspects of a situation to come into play. Focusing attention facilitates analytic thought by constraining activation such that items are considered in a compact form that is amenable to complex mental operations.

A plausible neural mechanism for CF has been proposed (Gabora, 2010; Gabora & Ranjan, 2013). In a state of defocused attention more aspects of a situation are processed, the set of activated properties is larger, and thus the set of possible associations is larger. Activation flows from specific instances, to the abstractions they instantiate, to other seemingly unrelated instances of those abstractions. Cell assemblies that would not be activated in analytic thought but that would be in associative thought are referred to as *neurds* (see also Ellamil, Dobson,

Beeman, & Christoff, 2012; Yoruk & Runco, 2014). Neurds respond to properties that are of marginal relevance to the current thought. They do not reside in any particular region of memory; the subset of cell assemblies that count as neurds shifts depending on the situation. For each different perspective one takes on an idea, a different group of neurds participates.

Neurds may generally be excluded from activated cell assemblies, becoming active only when there is a need to break out of a rut. In associative thought, diffuse activation causes more cell assemblies to be recruited, including neurds, enabling one thought to stray far from the preceding one while retaining a thread of continuity. Thus, the associative network is not just penetrated deeply, but traversed quickly. There is greater potential for overlapping representations to be experienced as wholes, resulting in the uniting of previously disparate ideas or concepts. While the preparation phase of the creative process likely involves long-term change to how ideas are encoded in the neocortex, the merging of thoughts culminating in insight may involve recurrent connections in the hippocampus, particularly when the insight involves generalization and inference triggered by a particular recent experience— (Kumaran & McClelland, 2012). Research on the neuroscience of insight suggests that alpha band activity in the right occipital cortex causes neural inhibition of sensory inputs, which enhances the relative influence of internally derived ("non-sensory") inputs, and thus the forging of new connections (Kounios & Beeman, 2009, 2014). Following insight, the shift to an analytic mode of thought could be accomplished through decruitment of neurds. Findings that the right hemisphere tends to engage in coarser semantic coding and have wider neuronal input fields than the left (Jung-Beeman, 2005; Kounios & Beeman, 2014) suggest that the right hemisphere may predominate during associative thought while the left predominates during analytic thought. In short, the architecture of associative memory is conducive to self-supervised learning and achievement of a more coherent network of understandings, with creative output as an external manifestation of this process.

## A New Conception of Divergent and Convergent Thought

Having examined how creative connections are forged in associative memory, we are led to the suggestion that instead of characterizing convergent thought in terms of the generation of a single correct solution, it be characterized in terms of sticking to conventional contexts, such as the context *car* when thinking of the concept TIRE. Similarly, it is proposed that divergent thought be characterized not by the generation of multiple solutions, but by capitalizing on the capacity to re-view an idea from new, context-relevant perspectives, using shared properties of concepts as bridges, to hone as few as a single idea. This is consistent with the RAT having only one solution yet requiring a more creative kind of thought than answering the question 'what is red', which has many possible solutions. Answering the question 'what is red' does not require thinking of concepts in unconventional contexts. In contrast, the RAT does require thinking of concepts in unconventional contexts. For example, consider the RAT question: what is the common associate of TANK, HILL and SECRET? The correct answer is TOP. To arrive at the correct answer, it is necessary to conceive of the concept TOP in very different ways.

For an everyday example of how this might work outside of creative thinking tests, consider how the concept TIRE might be brought to mind in a divergent mode of thought, as illustrated schematically in Figure 5. In its conventional context *car*, the concept TIRE collapses on tire-relevant properties such as 'goes on wheel' and 'filled with air'. However, in the unconventional context *playground equipment*, the concept TIRE might collapse on the properties that you could hang it and sit on it, which are essential for conceiving of it as a possible swing. In an even less conventional context, *pet needs*, the concept TIRE might collapses on the property 'small animal could sleep in it', which might enable it to be conceived of as a bed for a dog.

--------------------------------

Insert Figure 5 About Here

--------------------------------

This view of convergent and divergent thought arose as part of the honing theory of creativity (Gabora, in press), according to which creativity is a byproduct of the self-organizing, self-mending nature of a mind, and its proclivity to minimize what Hirsh, Mar, and Peterson (2012) refer to as psychological entropy: arousal-generating uncertainty. Honing theory grew out of a mathematical theory of concepts and their combinations, referred to as the State COntext Property theory (SCOP) (e.g., Aerts, Gabora, & Sozzo, 2013; Gabora & Aerts, 2002).As such, it takes seriously the need to formally model and study the chameleon-like way concepts change in response to new contexts, as well as the ill-formed, intermediate,or 'half-baked' states an idea can be in as it is being mulled over (e.g., Gabora & Carbert, 2015). Using data from a study in which participants rated the typicality of exemplars of a concept for different contexts, SCOP was able to model how the typicality of different contexts changes during a shift from a convergent (analytic) to a divergent (associative) mode of thought, such that seemingly atypical exemplars of the concept (e.g., PILON as an exemplar of the concept HAT) can emerge (Veloz, Gabora, Eyjolfson, & Aerts, 2011).

This conception of convergent and divergent thought is consistent with the widely-held view that divergent thought is conducive to insight, abduction, viewing situations from new perspectives, escaping fixation, and insight, while convergent thought is conducive to logic and the refinement of ideas. On the face of it, it is not obvious that fluency, or the capacity to generate many solutions, should equate with the kind of deep, prolonged, complex thought necessary for many creative accomplishments. However, if divergent thought is conceived of in terms of the capacity to conceive of the state of the problem or task in a new context, which yields a new state of the problem, and so on recursively until psychological entropy reaches an acceptable level, it becomes natural to equate divergent thinking with the kind of deep, prolonged, complex thought necessary for even big-C creativity.

### The Evolution of Convergent and Divergent Thought

Let us now examine how evolutionary considerations bear on the question of how divergent and convergent thinking problems are most fruitfully conceived. Does there exist a mechanism by which the brain could have evolved the capacity to engage in two kinds of thought processes, one for problems for which there is only one solution, and the other for problems that afford multiple solutions, as conventional views of convergent and divergent thinking would suggest? The answer must be no, for how could a brain even know, when the problem is first encountered, how many solution paths there are? Indeed, as strikingly demonstrated by the Indian mathematician Srinivasa Ramanujan, problems that are initially thought to have only one solution may later be revealed to be solvable by other means.

Now let us consider: does there exist a mechanism by which the brain would evolve the capacity to vary the extent to which context-specific aspects of a situation cause activation of the atypical yet potentially relevant properties that drive creative associations, as suggested by the view of convergent and divergent thinking proposed here? The answer is yes.

More specifically, a multifaceted program of research has been exploring the hypothesis that open-ended cultural evolution came about through two temporally-distinct cognitive transitions (Gabora, 2001, 2013; Gabora & Aerts, 2009; Gabora & Kaufman, 2010; Gabora & Kitto, 2013; Gabora & Smith, submitted). First, the emergence of Homo-specific culture approximately two million years ago, characterized most notably by the onset of primitive tool use, resulted from localized clustering of concepts, enabling the redescription and chaining of thoughts and actions, and the capacity for a stream of thought. This enabled a limited form of divergent thinking involving close associates but not remote ones.

Second, fully cognitive modernity and what Mithen (1996) refers to as the birth of art, science, and religion, following the appearance of anatomical modernity after 200,000 years ago during the Middle-Upper Paleolithic resulted from the onset of contextual focus (CF): the ability to shift along the spectrum from an extremely convergent mode on the one hand, to an extremely divergent mode on the other, involving remote associates as well as close ones. Thereafter, the fruits of divergent thought could now be used as ingredients for convergent thought, and vice versa. This paved the way for cognitive integration, which enabled the ongoing assimilation of new experiences and accommodation of the network of previous experiences.

It has been proposed that the onset of CF was made possible by a mutation of the FOXP2 gene known to have occurred in the Paleolithic period (Chrusch & Gabora, 2014). Although FOXP2 was initially called the "language gene", further research revealed that it is not uniquely associated with language. This suggests that the modified form of FOXP2 may have enabled the fine-tuning of the neurological mechanisms underlying the more general capacity to tailor their mode of thought to the situation at hand.

The proposal that the cultural transition of the Middle/Upper Paleolithic was due to the onset of contextual focus is consistent with Mithen's (1998) hypothesis that it was due to the onset of cognitive fluidity—the capacity to explore, map, and transform conceptual spaces across different knowledge systems—for contextual focus would enable one to engage in cognitive fluidity when it was appropriate, and then shift back to a more convergent mode of thought when unnecessary associations would be a distraction.

This two-transition theory is supported by simulations of chaining and CF carried out using an agent-based model of cultural evolution in which agents invent ideas for actions and imitate the fittest of their neighbors' actions (Gabora, Chia, & Firouzi, 2013). The mean fitness and diversity of actions across the model society increased with chaining, and even more so with CF, consistent with the hypothesis that these simulations broadly capture at an algorithmic level the mechanisms underlying the two cultural transitions. CF was particularly effective when the environment changed, which supports its hypothesized role in escaping fixation. CF has also been implemented in computational models of creativity, resulting in complex and fascinating works of music (Bell & Gabora, 2016) and art (DiPaola, & Gabora, 2009; McCaig, DiPaola, & Gabora, 2016).

<div align="center">Summary, Conclusions, and Practical Implications</div>

It has been proposed that by looking at the neural basis and evolution of creative thinking, it becomes apparent that creative thinking might fruitfully diverge, not in the sense that it (necessarily) results in the generation of multiple ideas, but in the sense of construing each concept in a manner that blurs distinctions between it and other concepts, thereby inviting associations. It is interesting that this is analogous to the way that 'divergent' is used with respect to light; divergent light does not consist of multiple small, focused beams, but rather a diffuse beam.

The proposed conception of divergent and convergent thinking suggests that conventional creativity tests as they are commonly used in educational and vocational settings may not be ideal. Both conventional tests of divergent thinking (such as the alternate uses test, which involves questions such as 'How many uses can you think of for a brick?) and of convergent thinking (such as the RAT discussed previously) may be assessing only a minor aspect of what the creative process entails. Creativity may be largely about tuning one's mode of thought in a context-specific manner such that each concept's halo of potential associations is tuned to match to the specifics of the task and how far along one is in it. To tap into this, it may be necessary to use a new breed of creativity tests that investigate how individuals shift between divergent and convergent modes of thought over the course of a creative task. Such tests are just beginning to be developed (e.g., Armen, 2015; Pringle, 2011; Pringle & Sowden, 2016; Pringle, Sowden, Deeley, & Sharma, in press).

## Acknowledgments

This work was supported by a grant (62R06523) from the Natural Sciences and Engineering Research Council of Canada.

References

Aerts, D., Gabora, L., & Sozzo, S. (2013). Concepts and their dynamics: A quantum-theoretical modeling of human thought. *Topics in Cognitive Science*, *5*, 737-772.

Armen, H. (2015). MetaCube: Using tangible interactions to shift between divergent & convergent thinking. *Proceedings of TEI*. Stanford CA: Association for Computing Machinery (ACM) Publications.

Bell, S. & Gabora, L. (2016). A music-generating system based on network theory. In *Proceedings of the 7th International Conference on Computational Creativity*. Palo Alto: Association for the Advancement of Artificial Intelligence (AAAI) Press.

Biggs, J. B., Fitzgerald, D., & Atkinson, S. M. (1971). Convergent and divergent abilities in children and teachers' ratings of competence and certain classroom behaviors. *British Journal of Educational Psychology, 41,* 277-286.

Carson, S., Peterson, J. B., & Higgins, D. M. (2003). Decreased latent inhibition is associated with increased creative achievement in high-functioning individuals. *Journal of Personality and Social Psychology, 85,* 499-506.

Chrusch, C. & Gabora, L. (2014). A tentative role for FOXP2 in the evolution of dual processing modes and generative abilities. In P. Bello, M. Guarini, M. McShane, & B. Scassellati (Eds.), *Proceedings of the 36th Annual Meeting of the Cognitive Science Society* (pp. 499-504). Austin TX: Cognitive Science Society.

DiPaola, S., & Gabora, L. (2009). Incorporating characteristics of human creativity into an evolutionary art algorithm. *Genetic Programming and Evolvable Machines, 10*, 97-110.

Ellamil, M. Dobson, C. Beeman, M. & Christoff, K. (2012). Evaluative and generative modes of thought during the creative process. *Neuroimage, 59,* 1783–1794.

Evans J. St. B. (2003). In two minds: dual process accounts of reasoning. *Trends in Cognitive Sciences, 7,* 454-59.

Gabora, L. (2001). *Cognitive mechanisms underlying the origin and evolution of culture.* Doctoral dissertation, Free University of Brussels.

Gabora, L. (2003). Contextual focus: A cognitive explanation for the cultural transition of the Middle/Upper Paleolithic. *Proceedings of the 25th annual meeting of the Cognitive Science Society* (pp. 432-437). Hillsdale NJ: Lawrence Erlbaum Associates.

Gabora, L. (2010). Revenge of the 'neurds': Characterizing creative thought in terms of the structure and dynamics of human memory. *Creativity Research Journal, 22*, 1-13.

Gabora, L. (2013). An evolutionary framework for culture: Selectionism versus communal exchange. *Physics of Life Reviews,10*, 117-145.

Gabora, L. (2014). Physical light as a metaphor for inner light. *Aisthesis, 7*, 43-61.Gabora, L. (2015). LIVEIA: A light-based immersive visualization environment for imaginative actualization: A new technology for psychological understanding. In S. Latifi (Ed.) *Proceedings of the 12th International Conference on Information Technology: New Generations* (pp. 686-691). Washington DC: IEEE Conference Publishing Services.

Gabora, L. (in press). Honing theory: A complex systems framework for creativity. *Nonlinear Dynamics, Psychology, and Life Sciences.*

Gabora, L., & Aerts, D. (2009). A model of the emergence and evolution of integrated worldviews. *Journal of Mathematical Psychology, 53,* 434-451.

Gabora, L. & Carbert, N. (2015). A study and preliminary model of cross-domain influences on creativity. In R. Dale, C. Jennings, P. Maglio, T. Matlock, D. Noelle, A. Warlaumont & J.

Yashimi (Eds.), *Proceedings of the 37th annual meeting of the Cognitive Science Society* (pp. 758-763). Austin TX: Cognitive Science Society.

Gabora, L., Chia, W., & Firouzi, H. (2013). A computational model of two cognitive transitions underlying cultural evolution. In M. Knauff, M. Pauen, N. Sebanz, & I. Wachsmuth (Eds.) *Proceedings of the 35th Annual Meeting of the Cognitive Science Society* (pp. 2344-2349). Austin TX: Cognitive Science Society.

Gabora, L., & Kaufman, S. (2010). Evolutionary perspectives on creativity. In J. C. Kaufman & R. J. Sternberg (Eds.), *The Cambridge handbook of Creativity* (pp. 279-300). Cambridge UK: Cambridge University Press.

Gabora, L., & Kitto, K. (2013). Concept combination and the origins of complex cognition. In (E. Swan, Ed.) *Origins of mind*: *Biosemiotics series, Vol. 8* (pp. 361-382). Berlin: Springer.

Gabora, L., & Ranjan, A. (2013). How insight emerges in distributed, content-addressable memory. In O. Vartanian, A. S. Bristol, & J. C. Kaufman (Eds.) *The neuroscience of creativity* (pp. 19-43). Cambridge, MA: MIT Press.

Guastello, S. J. (1998). Creative problem solving at the edge of chaos. *Journal of Creative Behavior 32*, 38-57.

Guilford, J. P. (1950). Creativity. *American Psychol*ogist, *5*, 444–454.

Hebb, D. (1949). *The organization of behavior*. New York: Wiley.

Hirsh, J. B., Mar, R. A., & Peterson, J. B. (2012). Psychological entropy: A framework for understanding uncertainty-related anxiety. *Psychological Review, 119*, 304-320.

Jung-Beeman M. (2005). Bilateral brain processes for comprehending natural language. *Trends in Cognitive Sciences, 9,* 512–518.

Kounios, J., & Beeman, M. (2009). The aha! moment: The cognitive neuroscience of insight. *Psychological Science*, *18*, 210–216. article.

Kounios, J., & Beeman, M. (2014). The cognitive neuroscience of insight. *Annual Review of Psychology*, *65*, 71–93.

Kumaran, D., & McClelland, J. (2012). Generalization through the recurrent interaction of episodic memories: A model of the hippocampal system. *Psychological Review, 119*, 573-616.

McCaig, G, DiPaola, S., & Gabora, L. (2016). Deep convolutional networks as models of generalization and blending within visual creativity. In *Proceedings of the 7th International Conference on Computational Creativity*. Palo Alto: Association for the Advancement of Artificial Intelligence (AAAI) Press.

Mithen, S. (1998). *Creativity in human evolution and prehistory*. Routledge.

Penn, D., Holyoak, K., & Povinelli, D. (2008). Darwin's mistake: Explaining the discontinuity between human and nonhuman minds. *Behavioral and Brain Sciences, 31,* 109-178.

Pringle, A. J. (2011). Shifting between modes of thought: A mechanism underlying creative performance? In *Proceedings of the 8th ACM conference on Creativity and cognition* (pp. 467-468). ACM.Pringle, A., & Sowden, P. T., (2016). The Mode Shifting Index (MSI): A new measure of the creative thinking skill of shifting between associative and analytic thinking. *Thinking Skills and Creativity*. DOI: 10.1016/j.tsc.2016.10.010

Pringle, A., Sowden, P. T., Deeley, C., & Sharma, S. (in press). Shifting between modes of thought: A domain-general creative thinking skill? *KIE Conference Publications*. Reprinted in In F. K. Reisman (Ed.) *Creativity in arts, science and technology*.

Riddell, W., Constans, E., Courtney, J., Dahm, K., Harvey, R., Jansson, P., Simone, M., & von Lockette, P. (2007). Lessons learned from teaching project based learning communication and design courses. *Proceedings of the 2007 Middle Atlantic Section Fall Conference of the American Society for Engineering Education*.

Runco, M. (2010). Divergent thinking, creativity, and ideation. In J. C. Kaufman & R. J. Sternberg, (Eds.), *The Cambridge handbook of creativity* (pp. 414-446). Cambridge UK: Cambridge University Press.

Sowden, P., Pringle, A., & Gabora, L. (2015). The shifting sands of creative thinking: Connections to dual process theory. *Thinking & Reasoning*, *21,* 40-60.

Veloz, T., Gabora, L., Eyjolfson, M., & Aerts, D. (2011). Toward a formal model of the shifting relationship between concepts and contexts in different modes of thought. In D. Song, M. Melucci, I. Frommholz, P. Zhang, L. Wang, & S. Arafat (Eds.), *Lecture notes in computer science 7052: Proceedings of the Fifth International Symposium on Quantum Interaction* (pp. 25-34). Berlin: Springer.

Yoruk S., & Runco M. A. (2014). The neuroscience of divergent thinking. *Activitas Nervosa Superior, 56*, 1–16.

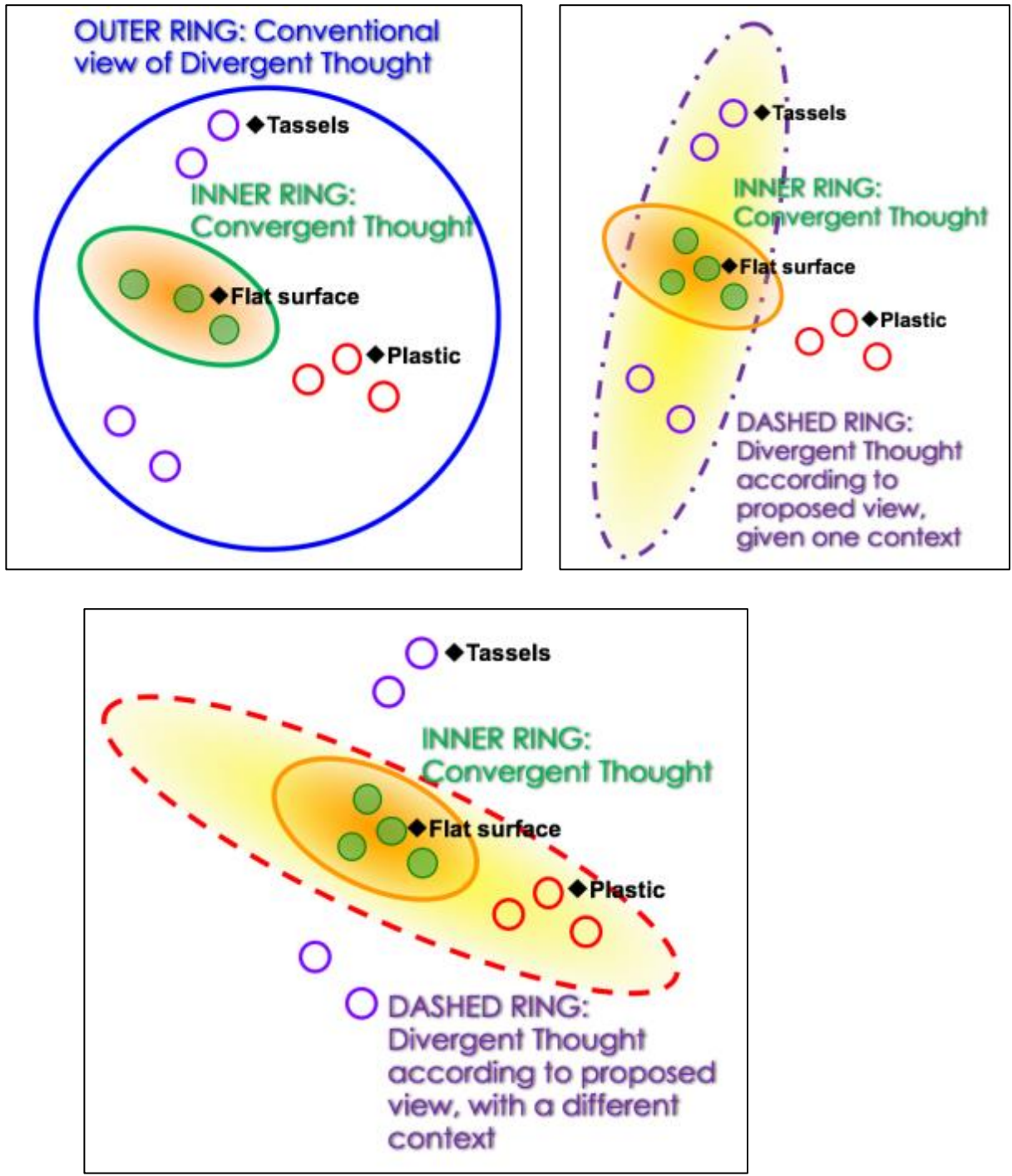

Figure 1. Neural-level illustration of context-dependency of creative thought. Small closed circles represent neural cell assemblies activated when thinking of CHAIR (regardless of which mode of thought). Small open circles represent neurons activated only in divergent thought referred to as 'neurds'. The top figure illustrates divergent thought as it is conventionally

construed. The bottom two illustrate the new conception of divergent thought. Notice that the subset of neurons that act as neurds in one context is not the same as the subset of neurons that act as neurds in a different context. Thus, given the context *living room*, the property 'tassels' might be activated (middle figure), but not in the context *outdoors*; in that context 'plastic' might be activated (bottom figure).

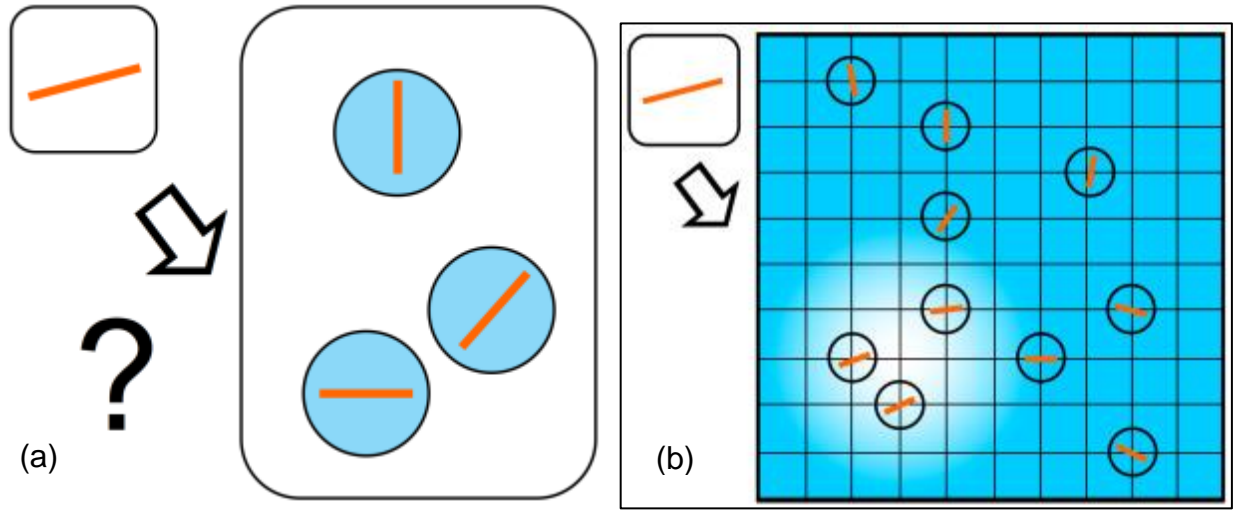

Figure 2. (a) In this schematic illustration of a portion of memory, the circles represent neurons, and the orange bars represent properties responded to by particular neurons—in this case, lines of a particular orientation. Since the total number of neurons in the brain is smaller than the total number of stimuli in the world that could potentially be encoded in memory, there exist stimuli that no neuron is tuned to respond to—such as, in this illustration, the line oriented at 15° to the left. Therefore, the question arises how the brain is nevertheless able to encode (and respond to) so many stimuli. (b) In this more detailed schematic representation of this portion of memory, each vertex represents a *possible* property, and each black ring represents a property that actually elicits maximal response from an existing neuron. (Thus, the fact that the number of neurons is smaller than the number of potential stimuli is represented by the fact that not all vertices have black circles.) The reason the 15° line can be encoded in memory is because its representation is *distributed*, or spread out across a cell assembly containing many neurons. The diffuse white circle indicates the region activated by the 15° line. For simplicity it contains only three neurons; in a real brain it would contain many more.

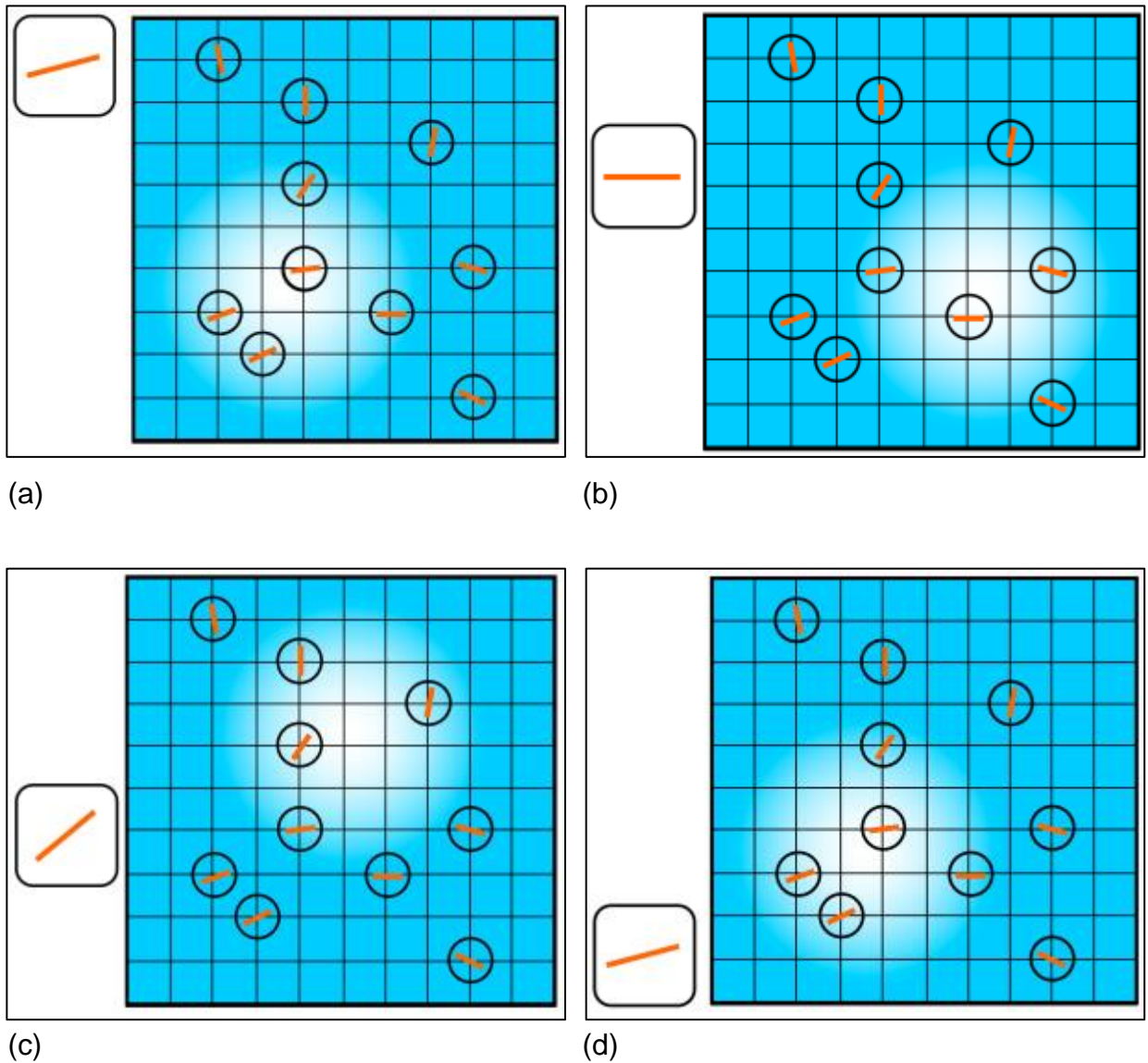

Figure 3. As in Figure 2, in this schematic illustration of a portion of memory, the circles represent neurons, and the orange bars represent properties responded to by particular neurons— in this case, lines of a particular orientation. Each of the four panels depicts a line of a particular orientation, and the corresponding region of memory activated by that stimulus. The fact that memory is *content-addressable* is illustrated by the fact that there is a systematic relationship between the stimulus *content* and *where* it gets encoded. Specifically, (b) depicts a stimulus that is similar to that in (a), and a nearby region of memory is activated, whereas (c) depicts a stimulus that is quite different to that in (a), and the region activated is further away. The stimulus in (d) is so similar to that in (a) that it activates the same region of memory.

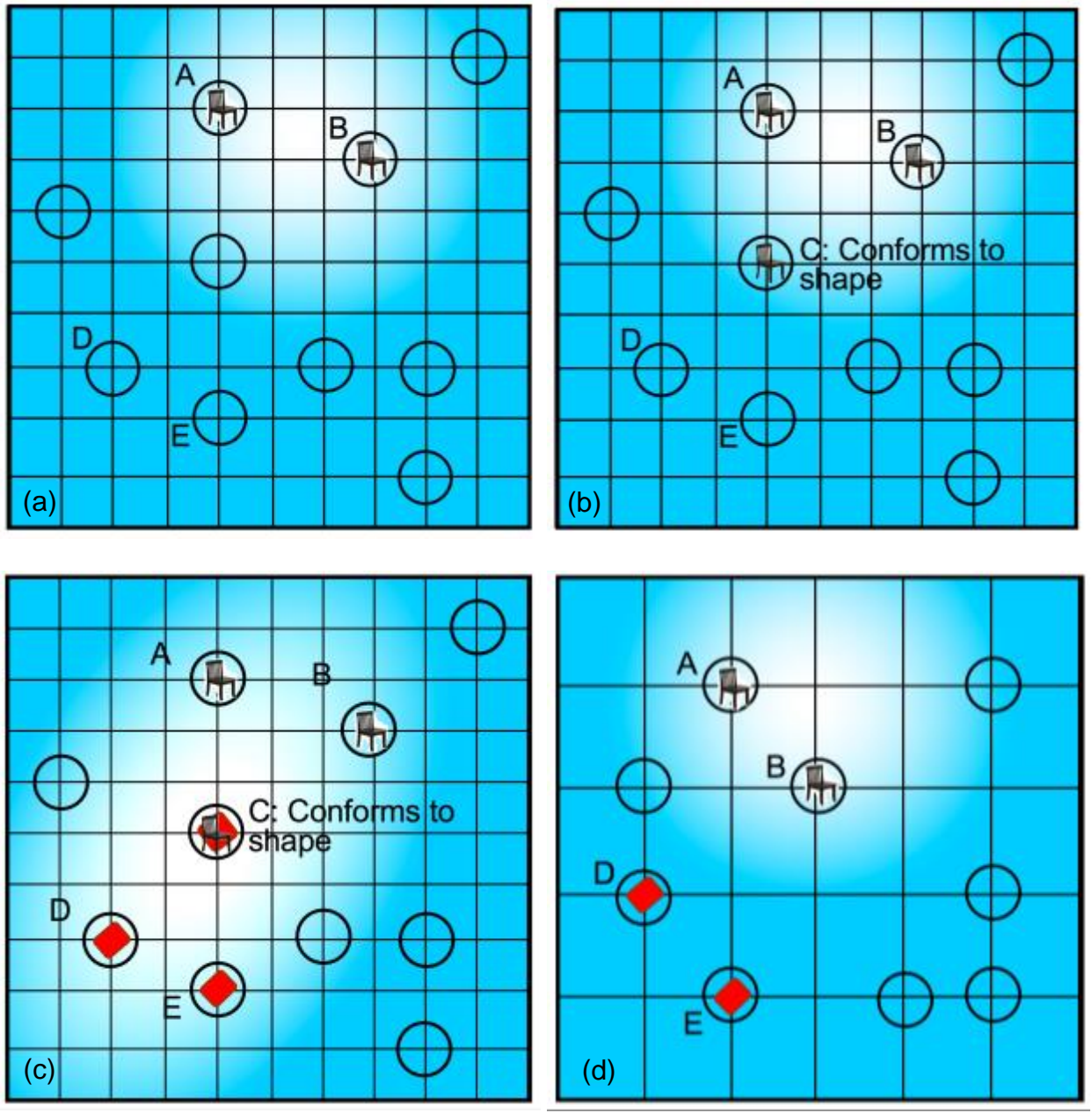

Figure 4. These panels provide a schematic illustration of a portion of memory in the process of inventing a beanbag chair. The circles represent neurons that respond to properties of objects such as chairs and beanbags. (a) The first panel depicts activation of the concept CHAIR in the absence of the goal of inventing a comfortable chair. Neurons that respond to only the most typical properties of CHAIR—such as that it has a flat surface and legs—are activated, and these are represented for simplicity as neurons A and B. (b) In the context of wanting to invent a *comfortable* chair, a fine-grained associative memory can expand its conception of CHAIR to additionally activate neurons that respond to context-relevant properties such as comfort. This is depicted as activation of neuron C that responds to 'conforms to shape'. (c) The concept BEANBAG is encoded in neurons C, D, and E, where C still responds to 'conforms to shape', and D and E respond to other typical beanbag properties such as that it is small and square. The overlap in the distributed representations of 'CHAIR in the context *comfortable*' and 'BEANBAG' means that there is a route by which the first can elicit associative retrieval of the second, culminating in invention of BEANBAG CHAIR. (d) The bottom right panel depicts an associative memory that encodes items in less detail. It does not contain a neuron that responds to objects with the property 'conforms to shape'. With CHAIR now activating only neurons A and B, and BEANBAG activating only neurons D and E, the context 'comfortable' cannot elicit associative retrieval of BEANBAG. Thus, this degree of detail is insufficient to bring about the invention of BEANBAG CHAIR.

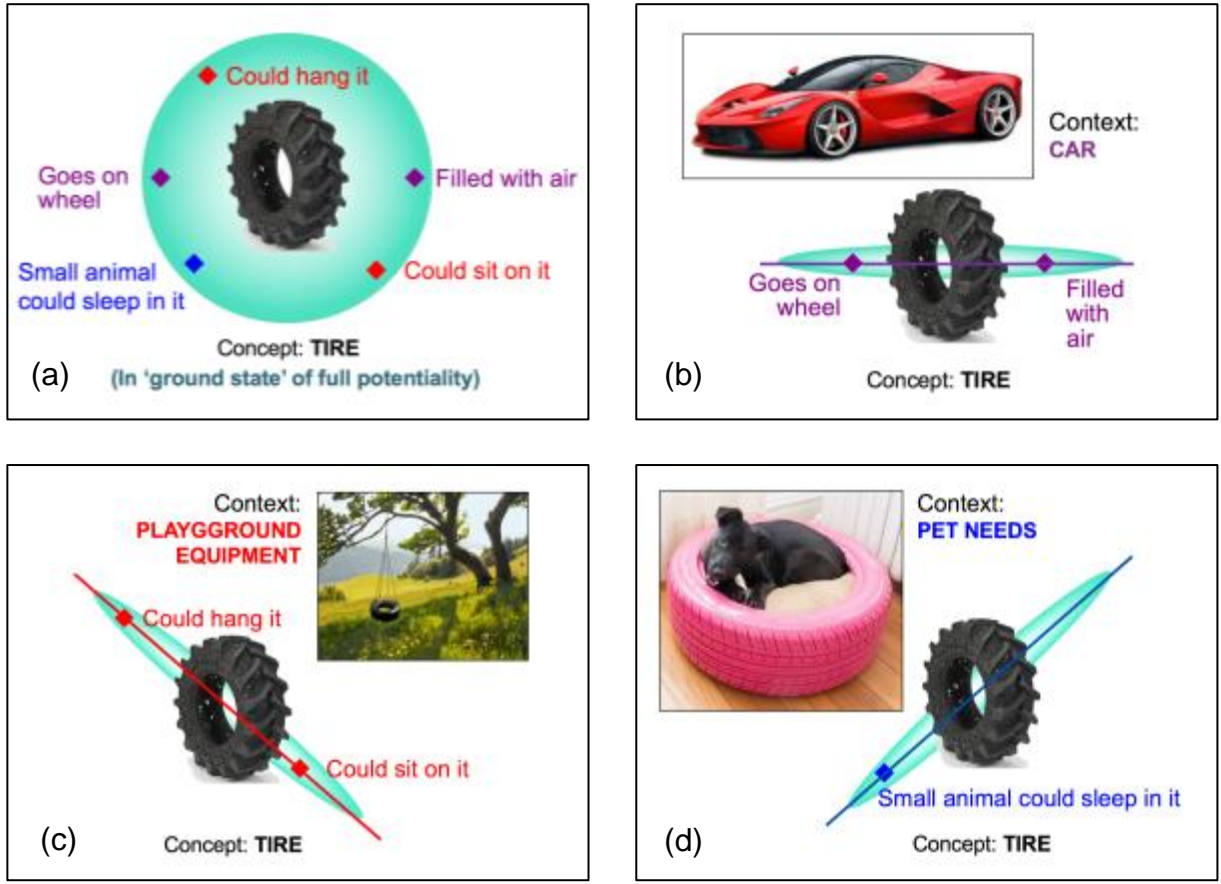

Figure 5. (a) The top left figure is a schematic depiction of the concept TIRE in its state of full potentiality, with many potential properties or affordances. (b) The top right figure depicts how, in its conventional context *car*, the concept TIRE collapses on tire-relevant properties such as 'goes on wheel' and 'filled with air'. (c) The bottom left figure depicts how, in the unconventional context *playground equipment*, the concept TIRE collapses on the properties that you could hang it and sit on it, which are essential for conceiving of it as a possible swing. (d) The bottom right figure depicts how, in an even more unconventional context for this concept, *pet needs*, it collapses on the property 'small animal could sleep in it', which is essential for conceiving of it as a dog bed.

---

[1] For more on how light can be used as a metaphor for cognition see (Gabora, 2014, 2015).

---

[2]  The earliest artifact that could be called a beanbag chair, referred to as a *sacco*, was

introduced in 1968 by three Italian designers: Piero Gatti, Cesare Paolini and Franco Teodoro

as part of the Italian Modernism movement.